\documentstyle[preprint,pre,aps,psfig,subfigure]{revtex}
\tightenlines

\begin{document}

\draft

\title{Error-correcting codes and image restoration\\
with multiple stages of dynamics}
\author{K. Y. Michael Wong}
\address{Department of Physics, Hong Kong University of Science and Technology,
Clear Water Bay, Kowloon, Hong Kong}
\author{Hidetoshi Nishimori}
\address{Department of Physics, Tokyo Institute of Technology,
Oh-Okayama, Meguro-ku, Tokyo 152-8551, Japan}

\date{\today}

\maketitle

\begin{abstract}
We consider the problems of error-correcting codes and image restoration 
with multiple stages of dynamics. 
Information extracted from the former stage
can be used selectively to improve the performance of the latter one.
Analytic results were derived for the mean-field systems 
using the cavity method. 
We find that it has the advantage of being tolerant
to uncertainties in hyperparameter estimation, 
as confirmed by simulations.
\end{abstract}


\narrowtext
\section{INTRODUCTION}

The corruption of signals by noise
is a common problem encountered in information processing.
To retrieve signals from messages
corrupted during the transmission through noisy channels,
various error-correcting codes have been proposed \cite{eliece}.
In particular, the error-correction mechanism
of a class of parity-checking codes
can be considered as the search for thermodynamically stable states
of a Hamiltonian constructed in terms of the message bits \cite{sourlas}.
These codes have been demonstrated
to saturate the Shannon information bound
in the limit that each encoded bit
checks the parity of an infinitely large number of message bits
\cite{sourlas,kabasaad}.
While in practice, each encoded bit can only check
the parity of a finite number of message bits,
these codes still maintain a very low bit error probability.

The need to retrieve signals from corrupted messages
is also inherent in image restoration \cite{geman}.
Although parity-checking bits
may not be explicitly introduced for the task,
prior knowledge about the images plays a similar role.
For example, the smoothness of real-world images
provides a mechanism for checking the pixel values
in comparison with those of their neighbors.
A corresponding Hamiltonian,
consisting of a ferromagnetic bias
to reflect the smoothening tendency,
can be constructed in terms of the image pixels.
Modern techniques of image restoration
based on Markov random fields correspond to
the search for thermodynamically stable states of the Hamiltonian system,
using methods such as simulated annealing \cite{geman}.

In a recent paper, we have shown that
the problems of error-correcting codes and image restoration
can be formulated in a unified framework \cite{nishiwong}.
In both tasks, the choice of the so-called hyperparameters
is an important factor in determining their performances.
Hyperparameters refer to the coefficients of the various interactions
appearing in the Hamiltonian of the tasks.
In error correction, they determine the statistical significance
given to the parity-checking terms and the received bits.
Similarly in image restoration,
they determine the statistical weights
given to the prior knowledge and the received data.
It was shown, by the use of inequalities,
that the optimal choice of the hyperparameters
correspond to the Maximum Posterior Marginal (MPM) method,
where there is a match between the source and model priors.
The choice of these values correspond to the Nishimori point
in the space of hyperparameters \cite{nishimori}.
It is equivalent to a thermodynamic process at finite temperature,
and the task performance is better than
the Maximum A Posteriori probability (MAP) method,
where the values of the hyperparameters are taken to infinity,
equivalent to a zero temperature process.
Furthermore, from the analytic solution of the infinite-range model
and the Monte Carlo simulation of finite-dimensional models,
it was shown that an inappropriate choice of the hyperparameters
can lead to a rapid degradation of the tasks.

In fact, hyperparameter estimation
has been the subject of many previous studies \cite{zhou},
a recently popular one using the ``evidence framework'' \cite{mackay}.
However, if the prior models the source poorly,
no hyperparameters can be reliable \cite{pryce}.
Even if they can be estimated accurately 
through steady-state statistical measurements, 
they may fluctuate when interfered 
by bursty noise sources in communication channels.
Hence it is important to devise decoding or restoration procedures
which are robust against the uncertainties in hyperparameter estimation.

In this paper we propose the technique of selective freezing
as a method to increase the tolerance to uncertainties
in hyperparameter estimation.
The technique has been studied 
for pattern reconstruction in neural networks,
where it led to an improvement in the retrieval precision,
a widening of the basin of attraction,
and a boost in the storage capacity \cite{wong}.
The idea is best illustrated
for Ising bits or pixels with binary states $\pm 1$,
though it can be easily generalized to other cases.
In a finite temperature thermodynamic process,
the Ising variables keep moving under thermal agitation.
Some of them have smaller thermal fluctuations than the others,
implying that they are more certain to stay in one state than the other.
This stability implies that they have a higher probability to stay
in the correct state for error-correction or image restoration tasks,
even when the hyperparameters are not optimally tuned.
It may thus be interesting to separate
the thermodynamic process into two stages.
In the first stage we select those relatively stable bits or pixels
whose time-averaged states have a magnitude exceeding a certain threshold.
In the second stage we subsequently fix (or freeze) them
in the most probable thermodynamic states
(for Ising variables this corresponds
to the sign of the time-averaged state).
Thus these selectively frozen bits or pixels
are able to provide a more robust assistance
to the less stable bits or pixels
in their search for the most probable states.
The selective freezing procedure
reduces to the usual finite-temperature decoding or restoration process
if all bits or pixels are frozen
(since nothing happens in the second stage),
or no bits or pixels are frozen
(since the second stage is merely a continuation
of the equilibration process of the first stage).

The two-stage thermodynamic process can be studied analytically
in the mean-field model,
which provides a qualitative guide
to the behavior of more realistic cases of lower dimensions.
However, it is necessary to give a remark
about the theoretical approach.
That is, as far as we have tried,
the analytical solution has been inaccessible
by the more conventional replica method.
Rather, we have to use the cavity method
to obtain the equations for the order parameters.
In particular, the cavity method leads to the appearance
of a term called the trans-susceptibility,
which correctly describes the effects of the thermodynamics 
of the first stage on that of the second.

The paper is organized as follows.
In Section II we briefly review the formulation
of error-correcting codes and image restoration
in a unified framework.
In Sections III and IV, we consider the mean-field model
for error-correcting codes and image restoration respectively.
We derive the equations for the order parameters
of the two-stage thermodynamics using the cavity method,
and present numerical results
illustrating the robustness of selective freezing
against uncertainties in hyperparameter estimation.
We further demonstrate that even when the noise model
changes without the receiver/restoration agent realizing the change
(i.e. it makes a wrong estimation of the prior),
the task performance is still robust.
For the more realistic cases of lower dimensions,
simulation results illustrate the relevance  of the infinite-range model
in providing qualitative guidance.
The conclusion is given in Section V.

\section{FORMULATION}

Consider an information source which generates data
represented by a set of Ising spins $\{\xi_i\}$,
where $\xi_i=\pm 1$ and $i=1, \cdots, N$.
The data is generated according to the source prior $P_s(\{\xi_i\})$.
For error-correcting codes transmitting unbiased messages,
all sequences are equally probable and $P_s(\{\xi\})=2^{-N}$.
For images with smooth structures,
the prior consists of ferromagnetic Boltzmann factors,
which increase the tendencies of the neighboring spins
to stay at the same spin states, that is,
\begin{equation}
        P_s(\{\xi\})=\frac{1}{Z(\beta_s)}
        \exp \left({\beta_s\over z}\sum_{\langle ij\rangle}
        \xi_i \xi_j \right).
\label{prior}
\end{equation}
Here $\langle ij\rangle$ represents pairs of neighboring spins, 
$z$ is the valency of each site, 
and the partition function $Z(\beta_s)$ is given by
\begin{equation}
        Z(\beta_s)={\rm Tr}_\xi
        \exp \left({\beta_s\over z}\sum_{\langle ij\rangle}
        \xi_i \xi_j \right).
\label{partition}
\end{equation}
The data is coded by constructing the codewords,
which are the products of $p$ spins 
$J^0_{i_1 \cdots i_p}=\xi_{i_1}\cdots\xi_{i_p}$
for appropriately chosen sets of of indices $\{i_1,\cdots,i_p\}$,
the choice of which determines the type of code.
Each spin may appear in a number of $p$-spin codewords;
the number of times of appearance is called the valency $z_p$.
The Sourlas code \cite{sourlas}
is equivalent to the infinite-range model
in which all possible codewords of $p$ spins
are chosen from $N$ spins.
On the other hand, the Kabashima-Saad code \cite{kabasaad}
consists of combinations in which
each spin appears in a random pre-selection of $z_p$ codewords.
For conventional image restoration,
codewords with only $p=1$ are transmitted,
corresponding to the pixels in the image;
the inclusion of terms with $p>1$,
and their positive effects on restoring the original image,
have also been discussed in \cite{nishiwong}.
For simplicity, we restrict ourselves
to the case of a single non-vanishing value of $p$
with $p\ge 2$, and $p=1$.

When the signal is transmitted through a noisy channel,
the output consists of the sets 
$\{ J_{i_1\cdots i_p}\}$ and $\{\tau_i\}$,
which are the corrupted versions of $\{J^0_{i_1 \cdots i_p}\}$
and $\{\xi_i\}$ respectively.
In the binary symmetric channel,
the outputs $J_{i_1 \cdots i_p}$ are equal to
$\mp J^0_{i_1 \cdots i_p}$ with probabilities
$p_J$ and $1-p_J$ respectively,
and $\tau_i$ equal to $\mp\xi_i$ with probabilities
$p_\tau$ and $1-p_\tau$ respectively.
Thus
\begin{equation}
        P_{\rm out}(\{J\},\{\tau\}|\{\xi\})
        \propto\exp\left(
        \beta_J\sum J_{i_1 \cdots i_p}\xi_{i_1}\cdots\xi_{i_p}
        +\beta_\tau\sum\tau_i\xi_i
        \right),
\label{BSC}
\end{equation}
where
\begin{equation}
        \beta_J={1\over 2}\ln\frac{1-p_J}{p_J}
        \quad{\rm and}\quad
        \beta_\tau={1\over 2}\ln\frac{1-p_\tau}{p_\tau}.
\end{equation}
The first summation in the exponent of Eq. (\ref{BSC}) extends
over an appropriate set of the indices $(i_1, \cdots ,i_p)$.

The Gaussian channel is defined by,
for a given sequence $\{\xi_i\}$,
\begin{equation}
        P_{\rm out}(\{ J\},\{\tau \}|\{\xi\}) 
        \propto\exp\left(
        -\frac{1}{2J^2}\sum
        (J_{i_1 \cdots i_p}-J_0\xi_{i_1}\cdots\xi_{i_p})^2
        -\frac{1}{2\tau^2}\sum
        (\tau_i-a\xi_i)^2\right).
\label{gauss}
\end{equation}
$J_0$ and $a$ are the strengths of the signals
to be fed into the channel,
and $J^2$ and $\tau^2$ are the variances of the noise.
We note that by letting $\beta_J$ and $\beta_\tau$ to be
$J_0/J^2$ and $a/\tau^2$ respectively,
the input-dependent terms of Eq. (\ref{gauss}) reduce
to those of Eq. (\ref{BSC}),
which therefore can be regarded as the noise model
for both binary symmetric and Gaussian channels.

According to Bayesian statistics,
the posterior probability that the source sequence is $\{\sigma\}$,
given the outputs $\{ J\}$ and $\{\tau\}$, takes the form
\begin{equation}
        P(\{\sigma\}|\{ J\},\{\tau\})
        \propto P_{\rm out}(\{J\},\{\tau\}|\{\sigma\})P_s(\{\sigma\}).
\end{equation}
Using Eq. (\ref{BSC}) and (\ref{prior}), we have
\begin{equation}
        P(\{\sigma\}|\{ J\},\{\tau\})
        \propto\exp\left(
        \beta_J\sum J_{i_1\cdots i_p}\sigma_{i_1}\cdots\sigma_{i_p}
        +\beta_\tau\sum\tau_i\sigma_i
        +{\beta_s\over z}\sum_{\langle ij\rangle}\sigma_i\sigma_j
        \right).
\label{Bayes0}
\end{equation}
It often happens that the receiver at the end of the
noisy channel does not have precise information
on $\beta_J$, $\beta_\tau$ or $\beta_s$.
One then has to estimate these parameters.
If the receiver estimates $\beta$, $h$ and $\beta_m$ 
for $\beta_J$, $\beta_\tau$ and $\beta_s$ respectively,
then the mean of the posterior distribution of $\sigma_i$
is equal to the thermal average
\begin{equation}
        \langle\sigma_i\rangle
        ={{\rm Tr}\sigma_i e^{-H\{\sigma\}}
        \over{\rm Tr} e^{-H\{\sigma\}}},
\label{sigmai}
\end{equation}
where the Hamiltonian is given by
\begin{equation}
        H\{\sigma\}
        =-\beta\sum J_{i_1 \cdots i_p}\sigma_{i_1}\cdots\sigma_{i_p}
        -h\sum\tau_i\sigma_i
        -{\beta_m\over z}\sum_{\langle ij\rangle}\sigma_i\sigma_j.
\label{Hamiltonian}
\end{equation}
One then regards ${\rm sgn}\langle\sigma_i\rangle$ as the
$i$th bit of the decoded/restored information.

To reduce the sensitivity of the decoding/restoration process
to the uncertainties in parameter estimation,
we propose a two-stage process of selective freezing
instead of the one-stage thermodynamic process
implied by Eq. (\ref{sigmai}).
In the first stage the spins evolve thermodynamically
as prescribed in Eq. (\ref{sigmai}),
and the thermal averages $\langle\sigma_i\rangle$
of the spins are monitored.
We may relate $\langle\sigma_i\rangle$ to an effective field $H_i$
by $\langle\sigma_i\rangle=\tanh H_i$.
Spins with larger magnitudes of $\langle\sigma_i\rangle$
correspond to larger magnitudes of $H_i$.
They are more likely to agree with the correct message or image bit,
and are less likely to change signs
even when the hyperparameters vary.
Their relative stability can be used
to assist the less stable spins
to boost their robustness against hyperparameter uncertainties.
Hence we select those spins with $|\langle\sigma_i\rangle|$
exceeding a given threshold $\theta$,
and freeze them in the second stage of the thermodynamics.
The average of the spin $\tilde\sigma_i$ in the second stage
is then given by
\begin{equation}
        \langle\tilde\sigma_i\rangle
        ={{\rm Tr}\tilde\sigma_i
        \prod_j\left[\Theta\left(\langle\sigma_j\rangle^2-\theta^2\right)
        \delta_{\tilde\sigma_j,{\rm sgn}\langle\sigma_j\rangle}
        +\Theta\left(\theta^2-\langle\sigma_j\rangle^2\right)\right]
        e^{-\tilde H\{\tilde\sigma\}}
        \over{\rm Tr}
        \prod_j\left[\Theta\left(\langle\sigma_j\rangle^2-\theta^2\right)
        \delta_{\tilde\sigma_j,{\rm sgn}\langle\sigma_j\rangle}
        +\Theta\left(\theta^2-\langle\sigma_j\rangle^2\right)\right]
        e^{-\tilde H\{\tilde\sigma\}}},
\end{equation}
where $\Theta$ is the step function,
$\tilde H\{\tilde\sigma\}$ is the Hamiltonian for the second stage,
and has the same form as Eq. (\ref{Hamiltonian}) in the first stage.
To increase the flexibility in the process,
the parameters $\beta$, $h$ and $\beta_m$ can be replaced by
$\tilde\beta$, $\tilde h$ and $\tilde\beta_m$ respectively
in the second stage.
One then regards ${\rm sgn}\langle\tilde\sigma_i\rangle$
as the $i$th spin of the decoding/restoration process.

The most important quantity in selective freezing
is the overlap of the decoded/restored bit
${\rm sgn}\langle\tilde\sigma_i\rangle$
and the original bit $\xi_i$ averaged
over the output probability and the spin distribution.
This is given by
\begin{equation}
        M_{\rm sf}
        =\sum_\xi\prod\int dJ\prod\int d\tau
        P_s(\{\xi\})P_{\rm out}(\{J\},\{\tau\}|\{\xi\})
        \xi_i{\rm sgn}\langle\tilde\sigma_i\rangle.
\label{M}
\end{equation}
Following Appendix A of \cite{nishiwong},
we can prove the following inequality
\begin{equation}
        M_{\rm sf}
        \le M(\beta=\beta_J, h=\beta_\tau, \beta_m=\beta_s),
\label{bound1}
\end{equation}
where the right hand side is the overlap
of the {\it single-stage} dynamics
when the model parameters $\beta$, $h$ and $\beta_m$ 
match the source parameters $\beta_J$, $\beta_\tau$ and $\beta_s$
respectively.
Hence selective freezing cannot outperform the single-stage process
if the hyperparameters can be estimated precisely.
However, we remark that the purpose of selective freezing
is rather to provide a relatively stable performance
when the hyperparameters cannot be estimated precisely.
This cannot be revealed from the inequality,
but will be confirmed by the analytic and simulation results
in Sections III and IV.

\section{THE INFINITE-RANGE MODEL FOR ERROR-CORRECTING CODES}

Let us now suppose that the output of the transmission channel
consists of only the set of $p$-spin interactions
$\{J_{i_1\cdots i_p}\}$.
The Hamiltonian (\ref{Hamiltonian}) then becomes
\begin{equation}
        H\{\sigma\}
        =-\beta\sum_{i_1<\cdots<i_p}
        J_{i_1\cdots i_p}\sigma_{i_1}\cdots\sigma_{i_p},
\label{p-hamiltonian}
\end{equation}
where we have set $\beta_m=0$
for the case that all messages are equally probable.

Analytical solutions for the overlap are in general unavailable.
We therefore consider the infinite-range model
in which the exchange interactions are present
for all possible pairs of sites
in the Hamiltonian of Eq. (\ref{p-hamiltonian}).

To consider the transition between error-free and errored regimes,
we are interested in the noise model
in which $J_{i_1\cdots i_p}$ is Gaussian
with mean $p!j_0\xi_{i_1}\cdots\xi_{i_p}/N^{p-1}$
and variance $p!J^2/2N^{p-1}$.
Since all messages are equally probable,
we can apply a gauge transformation $\sigma_i\to\sigma_i\xi_i$
and $J_{i_1\cdots i_p}\to J_{i_1\cdots i_p}\xi_{i_1}\cdots\xi_{i_p}$ 
to (\ref{p-hamiltonian}),
and arrive at an equivalent $p$-spin model
with a ferromagnetic bias, where
\begin{equation}
        P(J_{i_1\cdots i_p})
        =\left(N^{p-1}\over\pi J^2p!\right)^{1/2}
        \exp\left[-{N^{p-1}\over J^2p!}
        \left(J_{i_1\cdots i_p}-{p!\over N^{p-1}}j_0\right)^2\right].
\label{jdis}
\end{equation}
The Nishimori point for this model is located at $\beta=2j_0/J^2$.

The infinite-range model is exactly solvable
using mean-field theoretical techniques for disordered systems
such as the replica or cavity method \cite{mezard}.
Here we use the cavity method because of
its more transparent physical interpretation,
and some obstacles encountered in the use of the replica method.

The cavity method uses a self-consistency argument 
to consider what happens
when a spin is added or removed from the system. 
The central quantity in this method is the {\it cavity field}, 
which is the local field of a spin
when it is added to the system,
assuming that the exchange couplings act only one-way
from the system to the new spin
(but not from the spin back to the system).
Since the exchange couplings feeding the new spin
have no correlations with the system, 
the cavity field becomes a Gaussian variable
in the limit of large valency.

\subsection{Average spin in the first stage}

We start with the so-called ``clustering property''
for mean-field systems \cite{mezard},
\begin{equation}
        \langle\sigma_{i_1}\cdots\sigma_{i_p}\rangle
        =\langle\sigma_{i_1}\rangle\cdots\langle\sigma_{i_p}\rangle,
\label{cluster}
\end{equation}
where $\langle\quad\rangle$ represents thermodynamic averages.
As shown in Appendix A, the clustering property enables us
to express the thermal averages of a spin
in terms of the cavity field, say, for spin 1,
\begin{equation}
        \langle\sigma_1\rangle
        =\tanh\beta h_1;\quad
        h_1=\sum_{1<j_2<\cdots<j_p}J_{1j_2\cdots j_p}
        \langle\sigma_{j_2}\rangle^{\backslash 1}\cdots
        \langle\sigma_{j_p}\rangle^{\backslash 1},
\label{cavity}
\end{equation}
where the superscript $\backslash 1$ denotes the thermal averages
for a Hamiltonian in which
$\sigma_1$ and the associated exchange interactions are absent,
but otherwise identical to Eq. (\ref{p-hamiltonian}).
Thus $h_1$ is the cavity field obeying a Gaussian distribution,
whose mean and variance are
$pj_0m^{p-1}$ and $pJ^2q^{p-1}/2$ respectively,
where $m$ and $q$
are the magnetization and Edwards-Anderson order parameter
respectively, given by
\begin{equation}
	m\equiv{1\over N}\sum_i\langle\sigma_i\rangle
        \quad{\rm and}\quad
        q\equiv{1\over N}\sum_i\langle\sigma_i\rangle^2.
\label{mq}
\end{equation}
It is convenient to write
\begin{equation}
	\beta h_i=\hat m+\sqrt{\hat q}u_i,
\end{equation}
where
\begin{equation}
        \hat m=p\beta j_0m^{p-1}
        \quad{\rm and}\quad
        \hat q={p\over 2}\beta^2J^2q^{p-1},
\label{mqhat}
\end{equation}
and $u_i$ is a Gaussian variable with mean 0 and variance 1.

\subsection{Order parameters in the first stage}

Applying self-consistently the cavity argument
to all terms in Eq. (\ref{mq}),
we can obtain self-consistent equations for $m$ and $q$:
\begin{eqnarray}
        m&=&\int Du\tanh G,
\label{morder}\\
        q&=&\int Du\tanh^2 G,
\label{qorder}
\end{eqnarray}
where $Du\equiv du e^{-u^2/2}/\sqrt{2\pi}$ is the Gaussian measure,
$G=\hat m+\sqrt{\hat q}u$.
The overlap for the one-stage decoding process is given by
\begin{equation}
        M\equiv{1\over N}\sum_i{\rm sgn}\langle\sigma_i\rangle
        ={\rm erf}{\hat m\over\sqrt{2\hat q}}.
\end{equation}
Now we consider selective freezing.
If we introduce a freezing threshold $\theta$
so that all spins with $\langle\sigma_i\rangle^2>\theta^2$ are frozen,
then the freezing fraction $f$ is given by
\begin{equation}
        f\equiv{1\over N}\sum_i\Theta\left(
        \langle\sigma_i\rangle^2-\theta^2\right)
        =1-{1\over 2}{\rm erf}{u_+\over\sqrt{2}}
        +{1\over 2}{\rm erf}{u_-\over\sqrt{2}},
\end{equation}
where $u_\pm=(\pm u_0-\hat m)/\sqrt{\hat q}$
with $\tanh u_0=\theta$.

\subsection{Average spin in the second stage}

Assuming that the spin $\tilde\sigma_1$ is dynamic in the second stage,
we can write
\begin{equation}
        H\{\tilde\sigma\}\approx
        H\{\tilde\sigma\}^{\backslash 1}
        -\tilde\beta\sum_{1<j_1\cdots<j_{p-1}}
        \tilde\sigma_1J_{1j_1\cdots j_{p-1}}
        \prod_{s=1}^{p-1}\left[
        \tilde\sigma_{j_s}
        \Theta\left(\theta^2-\langle\sigma_{j_s}\rangle^2\right)
        +{\rm sgn}\langle\sigma_{j_s}\rangle
        \Theta\left(\langle\sigma_{j_s}\rangle^2-\theta^2\right)
        \right],
\label{2-hamiltonian}
\end{equation}
where $H\{\tilde\sigma\}^{\backslash 1}$
is the Hamiltonian when spin 1 is completely removed from the system
in both stages of the thermodynamic process.
Removing spin 1 may cause the thermal averages of other spins
to adjust slightly in the first stage.
Hence some dynamic spins (with $\langle\sigma_k\rangle^2<\theta^2$)
may become frozen ones (with $\langle\sigma_k\rangle^2>\theta^2$)
and vice versa, so that strictly speaking,
further terms should be considered in Eq. (\ref{2-hamiltonian})
to account for these secondary effects.
For example, if spin $k$ is induced
to switch from dynamic to frozen (or vice versa) on removal of spin 1,
then the Taylor expansion of $H\{\tilde\sigma\}$
implies that an extra term
\begin{eqnarray}
	&&-\tilde\beta\left(
	{\rm sgn}\langle\sigma_k\rangle^{\backslash 1}
	-\tilde\sigma_k\right)\left[
        \delta(\langle\sigma_k\rangle^{\backslash 1}-\theta)
        -\delta(\langle\sigma_k\rangle^{\backslash 1}+\theta)\right]
        (\langle\sigma_k\rangle-\langle\sigma_k\rangle^{\backslash 1})
	\nonumber\\
        &&\sum_{1<j_1\cdots<j_{p-1}\ne k}
        J_{kj_1\cdots j_{p-1}}
        \prod_{s=1}^{p-1}\left\{
        \tilde\sigma_{j_s}\Theta\left[
        \theta^2-(\langle\sigma_{j_s}\rangle^{\backslash 1})^2\right]
        +{\rm sgn}\langle\sigma_{j_s}\rangle^{\backslash 1}\Theta\left[
        (\langle\sigma_{j_s}\rangle^{\backslash 1})^2-\theta^2\right]
        \right\}
\end{eqnarray}
should be incorporated in Eq. (\ref{2-hamiltonian}).
Here, we have neglected these terms for clarity.
Nevertheless, justification a posteriori
can be provided for their deletion.

Using a cavity argument similar to Appendix A,
we can show that 
\begin{equation}
        \langle\tilde\sigma_1\rangle
        =\tanh\tilde\beta\left\{\sum_{1<j_1\cdots<j_{p-1}}
        J_{1j_1\cdots j_{p-1}}
        \prod_{s=1}^{p-1}\left[
        \langle\tilde\sigma_{j_s}\rangle^{\backslash 1}
        \Theta\left(\theta^2-\langle\sigma_{j_s}\rangle^2\right)
        +{\rm sgn}\langle\sigma_{j_s}\rangle^{\backslash 1}
        \Theta\left(\langle\sigma_{j_s}\rangle^2-\theta^2\right)
        \right]\right\}.
\label{2-average}
\end{equation}
However, the effective field
on the right hand side of Eq. (\ref{2-average})
is still not a cavity field
because $\langle\sigma_{j_s}\rangle$,
which are used in the step functions to decide
whether the spin $j_s$ is dynamic or frozen in the second stage,
is different from $\langle\sigma_{j_s}\rangle^{\backslash 1}$.
Hence it may have correlations with spin 1.
Taylor expansion of $\langle\sigma_{j_s}\rangle$
about $\langle\sigma_{j_s}\rangle^{\backslash 1}$ yields
\begin{eqnarray}
        &&\langle\tilde\sigma_1\rangle
        =\tanh\tilde\beta\Biggl\{\tilde h_1
        +\sum_{1j\ne j_1\cdots<j_{p-2}}
        J_{1jj_1\cdots j_{p-2}}\nonumber\\
        &&\prod_{s=1}^{p-2}\left[
        \langle\tilde\sigma_{j_s}\rangle^{\backslash 1}
        \Theta\left[\theta^2
        -(\langle\sigma_{j_s}\rangle^{\backslash 1})^2\right]
        +{\rm sgn}\langle\sigma_{j_s}\rangle^{\backslash 1}
        \Theta\left[
        (\langle\sigma_{j_s}\rangle^{\backslash 1})^2-\theta^2\right]\right]
	\nonumber\\
        &&\left[{\rm sgn}\langle\sigma_j\rangle^{\backslash 1}
        -\langle\tilde\sigma_j\rangle^{\backslash 1}\right]
        \left[\delta(\langle\sigma_j\rangle^{\backslash 1}-\theta)
        -\delta(\langle\sigma_j\rangle^{\backslash 1}+\theta)\right]
        (\langle\sigma_j\rangle-\langle\sigma_j\rangle^{\backslash 1})
        \Biggr\},
\label{2-without1}
\end{eqnarray}
where $\tilde h_1$ is the generic cavity field
which is now completely uncorrelated with spin 1.
It is given by
\begin{equation}
        \tilde h_1
        =\sum_{1<j_1\cdots<j_{p-1}}
        J_{1j_1\cdots j_{p-1}}
        \prod_{s=1}^{p-1}\left\{
        \langle\tilde\sigma_{j_s}\rangle^{\backslash 1}
        \Theta\left[\theta^2
        -(\langle\sigma_{j_s}\rangle^{\backslash 1})^2\right]
        +{\rm sgn}\langle\sigma_{j_s}\rangle^{\backslash 1}
        \Theta\left[(\langle\sigma_{j_s}\rangle^{\backslash 1})^2
        -\theta^2\right]
        \right\}.
\label{2-cavity}
\end{equation}
To evaluate the difference
$\langle\sigma_j\rangle-\langle\sigma_j\rangle^{\backslash 1}$
appearing in Eq. (\ref{2-without1}),
we have to apply the cavity method a second time,
by comparing the changes when both spins 1 and $j$ are removed.
This is done in Appendix B and the result is
\begin{equation}
        \langle\sigma_j\rangle-\langle\sigma_j\rangle^{\backslash 1}
        =\left(\beta{\rm sech}^2\beta h_j^{\backslash 1}\right)
	\left(h_{j1}\tanh\beta h_1^{\backslash j}\right),
\label{2-change}
\end{equation}
where
\begin{equation}
        h_{1j}=h_{j1}
        =\sum_{1j\ne k_1\cdots<k_{p-2}}
        J_{1jk_1\cdots k_{p-2}}
        \langle\sigma_{k_1}\rangle^{\backslash 1j}\cdots
        \langle\sigma_{k_{p-2}}\rangle^{\backslash 1j}.
\label{h_1j}
\end{equation}
When Eqs. (\ref{2-change}-\ref{h_1j})
are substituted into Eq. (\ref{2-without1}),
the significant contribution comes from the terms
which pair up $J_{1jj_1\cdots j_{p-2}}$ and $J_{1jk_1\cdots k_{p-2}}$.
The various terms appearing in the summation over
$j\ne j_1<\cdots<j_{p-2}$
involve thermal averages in the absence of spins 1 or $j$.
We assume that the effects of removing a spin is negligible
(which can be shown to be equivalent
to the replica symmetric approximation
in the replica method \cite{wong2}).
Then replacing the components of the terms
by their mean values,
and counting that $N^{p-2}/(p-2)!$ terms
appearing in the summation over $j_1<\cdots<j_{p-2}$,
we arrive at
\begin{eqnarray}
        &&\langle\tilde\sigma_1\rangle
        =\tanh\tilde\beta\Biggl\{\tilde h_1
        +{p\over 2}(p-1) J^2 {1\over N}\sum_j
        \left[\delta(\langle\sigma_j\rangle-\theta)
        -\delta(\langle\sigma_j\rangle+\theta)\right]
        \left[{\rm sgn}\langle\sigma_j\rangle
        -\langle\tilde\sigma_j\rangle\right]\nonumber\\
        &&\left(\beta{\rm sech}^2\beta h_j\right)
	\left(r^{p-2}\tanh\beta h_1\right)\Biggr\},
\label{2-paired}
\end{eqnarray}
where $r$ is the order parameter
describing the spin correlations of the two thermodynamic stages:
\begin{equation}
        r\equiv
        {1\over N}\sum_i
        \langle\sigma_i\rangle\left\{
        \langle\tilde\sigma_i\rangle
        \Theta\left[\theta^2-\langle\sigma_i\rangle^2\right]
        +{\rm sgn}\langle\sigma_i\rangle
        \Theta\left[\langle\sigma_i\rangle^2-\theta^2\right]
        \right\}.
\label{r}
\end{equation}
Eq. (\ref{2-paired}) can be simplified
by introducing the trans-susceptibility $\chi_{tr}$,
which describes the response of a spin in the second stage
to variations of the cavity field in the first stage, namely
\begin{equation}
        \chi_{tr}
	\equiv{1\over N}\sum_i
        {\partial\langle\tilde\sigma_i\rangle\over
        \partial h_i}.
\end{equation}
Since $\langle\tilde\sigma_i\rangle$ equals ${\rm sgn}h_i$
for $\tanh^2\beta h_i>\theta^2$,
and $\tanh\beta\tilde h_i$ otherwise, we get
\begin{equation}
        \chi_{tr}
        ={1\over N}\sum_i
        \left[\delta(\langle\sigma_i\rangle-\theta)
        -\delta(\langle\sigma_i\rangle+\theta)\right]
        \left[{\rm sgn}\langle\sigma_i\rangle
        -\langle\tilde\sigma_i\rangle\right]
        \beta{\rm sech}^2\beta h_i.
\label{trans}
\end{equation}
Eq. (\ref{2-paired}) can thus be simplified to
\begin{equation}
        \langle\tilde\sigma_1\rangle
        =\tanh\tilde\beta\left\{\tilde h_1
        +{p\over 2}(p-1) J^2 r^{p-2} \chi_{tr}
        \tanh\beta h_1\right\}.
\label{2-result}
\end{equation}

\subsection{Order parameters in the second stage}

The cavity field $\tilde h_1$ in the second stage is a Gaussian variable.
Its mean and variance are 
$pj_0\tilde m^{p-1}$ and $pJ^2\tilde q^{p-1}/2$ respectively,
where $\tilde m$ and $\tilde q$
are the magnetization and Edwards-Anderson order parameter respectively,
given by
\begin{eqnarray}
        \tilde m
	&\equiv&{1\over N}\sum_i
        \left[\Theta(\theta^2-\langle\sigma_i\rangle^2)
        \langle\tilde\sigma_i\rangle
        +\Theta(\langle\sigma_i\rangle^2-\theta^2)
        {\rm sgn}\langle\sigma_i\rangle\right],
\label{2-m}\\
        \tilde q
	&\equiv&{1\over N}\sum_i
        \left[\Theta(\theta^2-\langle\sigma_i\rangle^2)
        \langle\tilde\sigma_i\rangle^2
        +\Theta(\langle\sigma_i\rangle^2-\theta^2)
        \right].
\label{2-q}
\end{eqnarray}
Furthermore, the covariance between $h_1$ and $\tilde h_1$
is $pJ^2 r^{p-1}/2$, where $r$ is given in Eq. (\ref{r}).

Algebraic manipulations can be simplified if we write, for $i=1$,
\begin{eqnarray}
        \beta h_i
        &=&\hat m+\sqrt{\hat q}u_i,\\
        \tilde\beta\tilde h_i
        &=&\hat{\tilde m}+\sqrt{\hat{\tilde q}}
        (\eta u_i+\sqrt{1-\eta^2}v_i),
\end{eqnarray}
where $u_i$ and $v_i$ are independent Gaussian variables
with mean 0 and variance 1,
$\hat m$, $\hat q$ are given in Eq. (\ref{mqhat}), and
\begin{eqnarray}
        &&\hat{\tilde m}=p\tilde\beta j_0\tilde m^{p-1},
        \quad{\rm and}\quad
        \hat{\tilde q}={p\over 2}\tilde\beta^2J^2\tilde q^{p-1},\\
        &&\hat r={p\over 2}\beta\tilde\beta J^2 r^{p-1},
        \quad{\rm and}\quad
        \eta={\hat r\over\sqrt{\hat q\hat{\tilde q}}}.
\label{2-mqhat}
\end{eqnarray}
Applying self-consistently the same cavity argument
to all terms in Eqs. (\ref{2-m}), (\ref{2-q}), (\ref{r}) and (\ref{trans})
and performing the Gaussian average over $u_i$ and $v_i$,
we arrive at the following self-consistent equations
for $\tilde m$, $\tilde q$, $r$ and $\chi_{tr}$:
\begin{eqnarray}
        \tilde m&=&
        -{1\over 2}{\rm erf}{u_+\over\sqrt 2}
        -{1\over 2}{\rm erf}{u_-\over\sqrt 2}
        +\int_{u_-}^{u_+}Du\int Dv\tanh L,
\label{2-morder}\\
        \tilde q&=&
        1-{1\over 2}{\rm erf}{u_+\over\sqrt 2}
        +{1\over 2}{\rm erf}{u_-\over\sqrt 2}
        +\int_{u_-}^{u_+}Du\int Dv\tanh^2 L,
\label{2-qorder}\\
        r&=&
        \left(\int_{-\infty}^{u_-}+\int_{u_+}^\infty\right)Du
        \left|\tanh G\right|
        +\int_{u_-}^{u_+}Du\int Dv\tanh G\tanh L,
\label{rorder}\\
        \chi_{tr}&=&
        {\exp(-u_+^2/2)\over J\sqrt{\pi pq^{p-1}}}
        \int Dv(1-\tanh L_v^{(+)})+
        {\exp(-u_-^2/2)\over J\sqrt{\pi pq^{p-1}}}
        \int Dv(1+\tanh L_v^{(-)}),
\label{transorder}
\end{eqnarray}
where
\begin{eqnarray}
        L&=&
        \hat{\tilde m}+\sqrt{\hat{\tilde q}}(\eta u+\sqrt{1-\eta^2}v)
        +{p\over 2}(p-1)\tilde\beta J^2 r^{p-2} \chi_{tr} \tanh G,\\
        L_v^{(\pm)}&=&
        \hat{\tilde m}+\sqrt{\hat{\tilde q}}(\eta u_\pm+\sqrt{1-\eta^2}v)
        \pm{p\over 2}(p-1)\tilde\beta J^2 r^{p-2} \chi_{tr} \theta.
\end{eqnarray}
Eqs. (\ref{morder}-\ref{qorder}), (\ref{2-morder}-\ref{transorder})
for the order parameters
$m$, $q$, $\tilde m$, $\tilde q$, $r$ and $\chi_{tr}$
form a close set of equations.
The performance of selective freezing is measured by
\begin{equation}
        M_{\rm sf}
        \equiv{1\over N}\sum_i
        \left[\Theta(\theta^2-\langle\sigma_i\rangle^2)
        {\rm sgn}\langle\tilde\sigma_i\rangle
        +\Theta(\langle\sigma_i\rangle^2-\theta^2)
        {\rm sgn}\langle\sigma_i\rangle\right].
\end{equation}
From the above parameters, $M_{\rm sf}$ can be derived as:
\begin{equation}
        M_{\rm sf}
        =-{1\over 2}{\rm erf}{u_+\over\sqrt 2}
        -{1\over 2}{\rm erf}{u_-\over\sqrt 2}
        +\int_{u_-}^{u_+}Du
        {\rm erf}{L_u\over\sqrt{2\tilde q(1-\eta^2)}},
\end{equation}
where $L_u=\hat{\tilde m}+\sqrt{\hat{\tilde q}}\eta u
+[p(p-1)/2]\tilde\beta J^2 r^{p-2} \chi_{tr} \tanh G$.

We have also tried to derive the above equations
using the replica method.
However, in the nearest results that we could find,
terms involving the trans-susceptbility are absent,
which we believe to be unphysical.
Therefore the replica approach to the order parameter equations
remain an open question.

\begin{figure}
\centering
\centerline{\psfig{figure=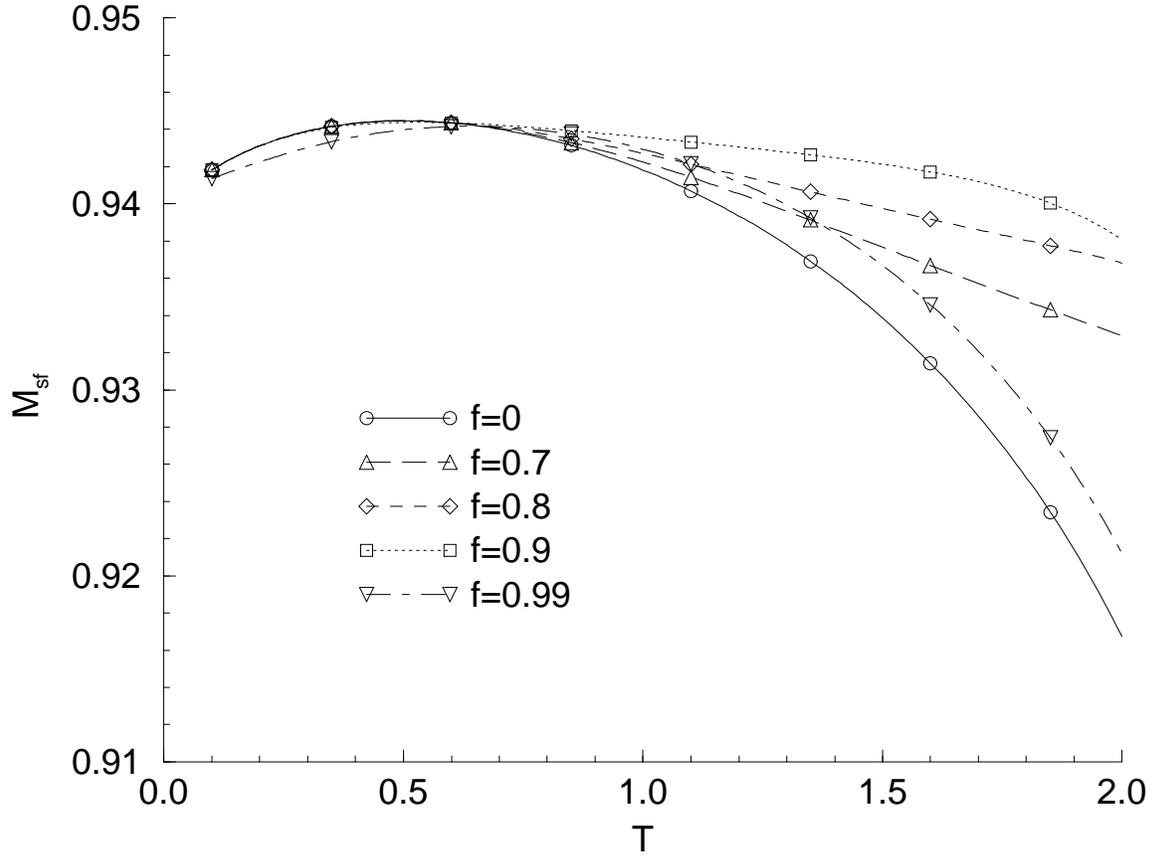,height=13cm}}
\caption{
The overlap $M_{\rm sf}$ as a function of the decoding temperature $T$ 
for $p=2$ and $j_0=J=1$ 
for various given values of freezing fraction $f$.
In this and the following figures,
$f=0$ corresponds to one-stage decoding/restoration.
}
\label{bif.vgr}
\end{figure}

We show an example of the case $p=2$ and $j_0=J=1$ in Fig. \ref{bif.vgr},
where the overlap $M_{\rm sf}$ is plotted as a function
of the decoding temperature $T (=\beta^{-1}=\tilde\beta^{-1})$
for various given values of freezing fraction $f$.
When $f=0$ (no spins frozen) and $f=1$ (all spins frozen),
the dynamics is equivalent to one with single stage,
and the overlap reaches its maximum
at the Nishimori point $T=J^2/2j_0$ as expected.
We observe that the tolerance against variations in $T$
is enhanced by selective freezing for certain values of $f$.

\begin{figure}
\centering
\centerline{\psfig{figure=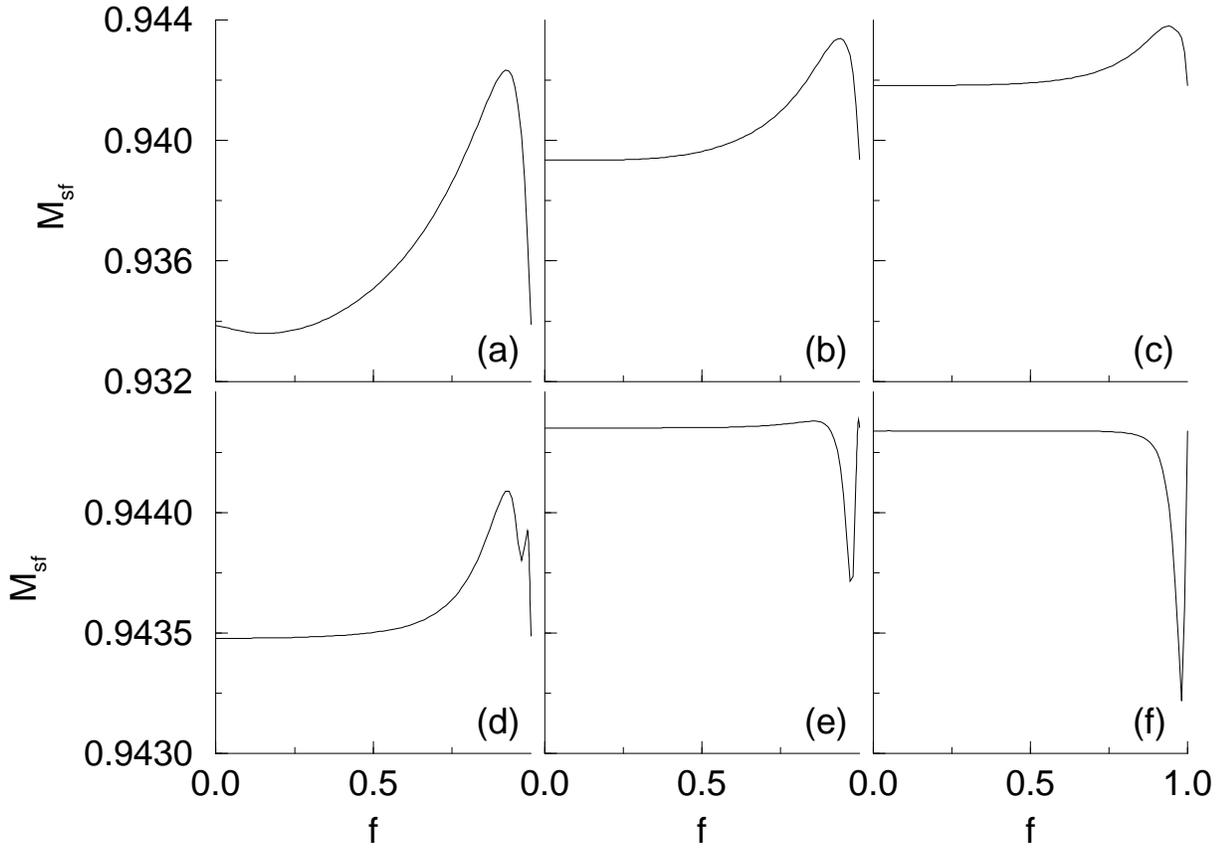,height=13cm}}
\caption{
The overlap $M_{\rm sf}$ as a function of the freezing fraction $f$ 
at temperatures $T$=(a) 1.5, (b) 1.2, (c) 1.0, (d) 0.8, (e) 0.6, (f) 0.4.
for $p=2$ and $j_0=J=1$.
}
\label{bif2.vgr}
\end{figure}

It is therefore interesting to consider the appropriate values of $f$
for the best overlap at a given decoding temperature.
Figs. \ref{bif2.vgr}(a-f) shows that at high temperatures 
such as in Figs. \ref{bif2.vgr}(a-c),
there is a single maximum 
and its position is fairly independent of temperature,
lying around $f=0.9$ in the present case.
At intermediate temperatures such as in Figs. \ref{bif2.vgr}(d-e),
there appear two maxima and as temperature changes,
there is a discontinuous jump in the maximum position.
Fig. \ref{bif2.vgr}(f) shows that when the temperature is lower
than the Nishimori point ($T_N=0.5$),
the overlap cannot be improved by selective freezing.

\begin{figure}
\centering
\centerline{\psfig{figure=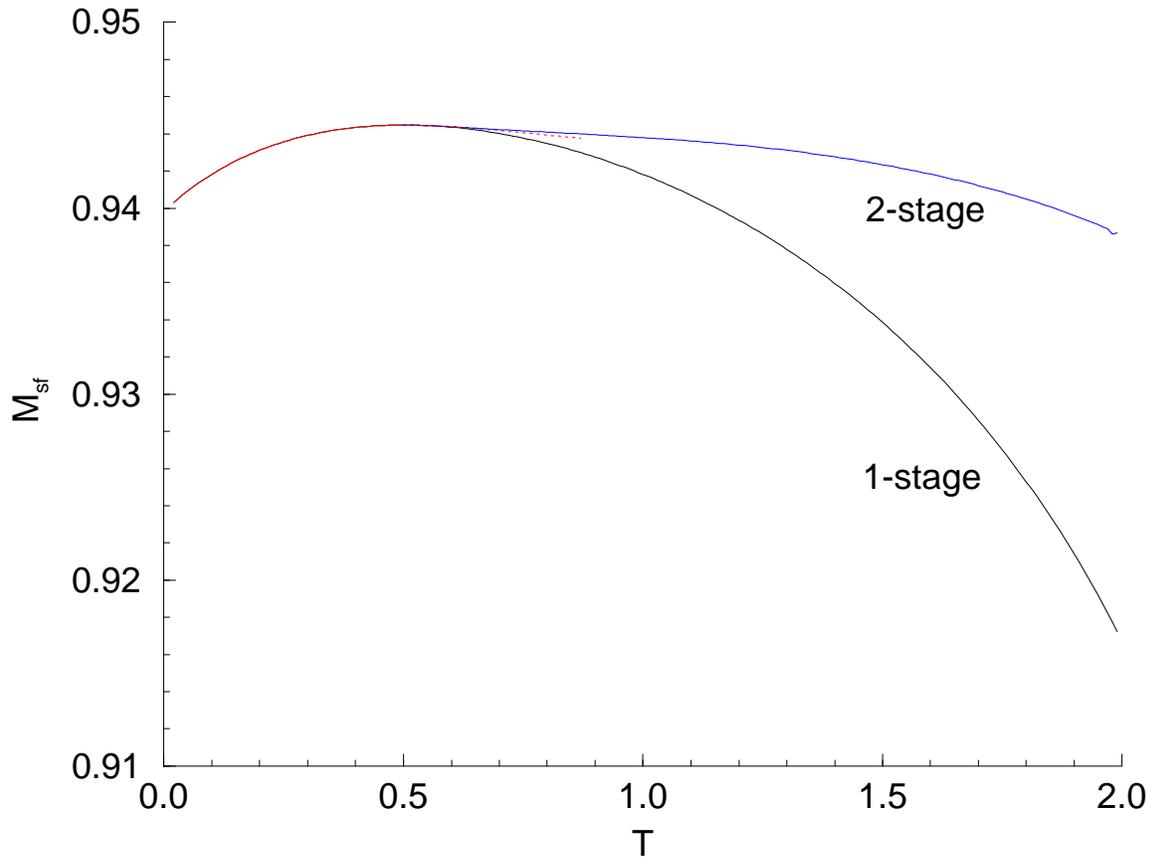,height=13cm}}
\caption{
The temperature dependence of the best overlap of selective freezing 
compared with the overlap of the one-stage dynamics 
for $p=2$ and $j_0=J=1$.
}
\label{bit3.vgr}
\end{figure}

\begin{figure}
\centering
\centerline{\psfig{figure=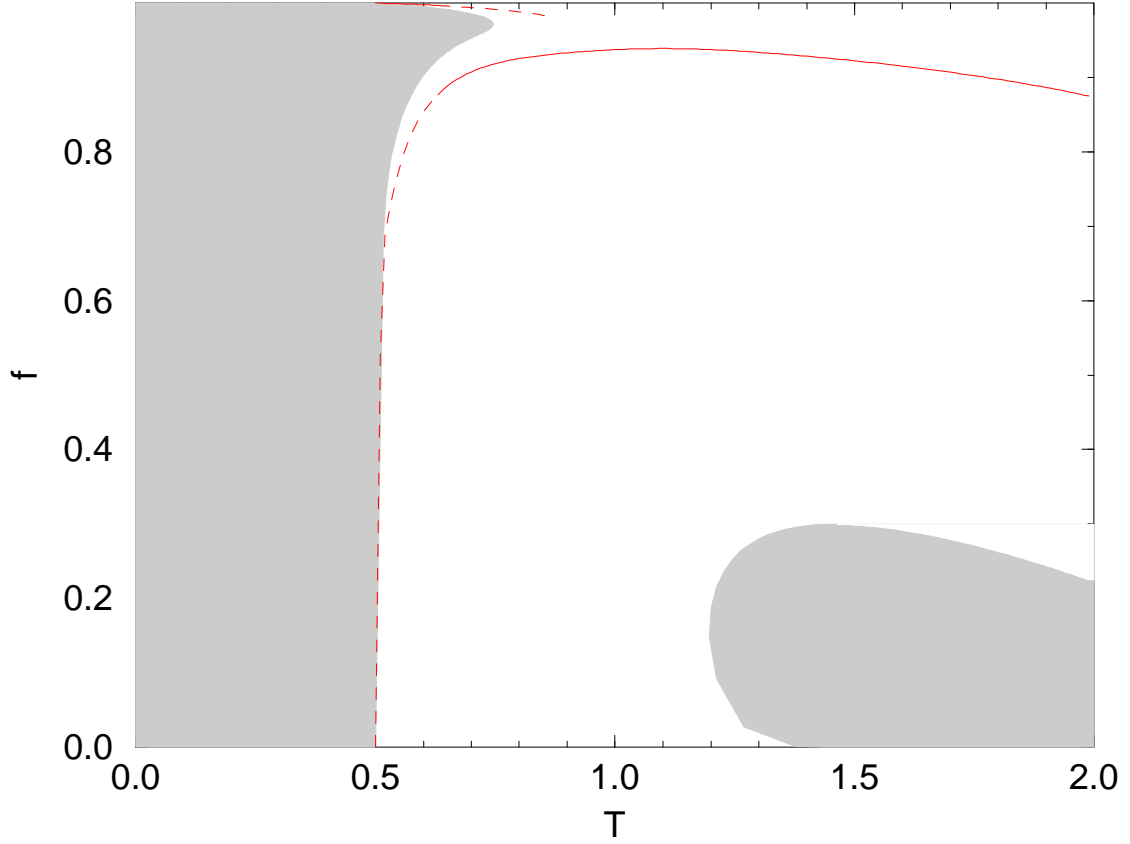,height=13cm}}
\caption{
The freezing fraction $f$ for the best overlap 
as a function of temperature $T$ 
for $p=2$ and $j_0=J=1$. 
In this and the following figure, 
solid line: global maximum, 
dashed line: local maximum, 
shaded region: $M_{\rm sf}<M$.
}
\label{bif4.vgr}
\end{figure}

Figure \ref{bit3.vgr} compares the overlap of the one-stage dynamics
with that of the best of selective freezing.
It shows that when the decoding temperature is mis-determined
to be higher than its optimal value at the Nishimori point,
selective freezing can provide a fairly robust performance.
Furthermore, the choice of the freezing fraction
for such robust performance
appears to be quite independent of the temperature.
The solid line in Fig. \ref{bif4.vgr} locates the position 
for the best overlap and, as observed from Figs. \ref{bif2.vgr}(a-f),
lies in the vicinity of $f\approx 0.9$ for a large range of temperature.
The unshaded region in the same figure
also indicates that selective freezing
leads to an improvement in the overlap
over a wide range of the parameter space.

\begin{figure}
\centering
\centerline{\psfig{figure=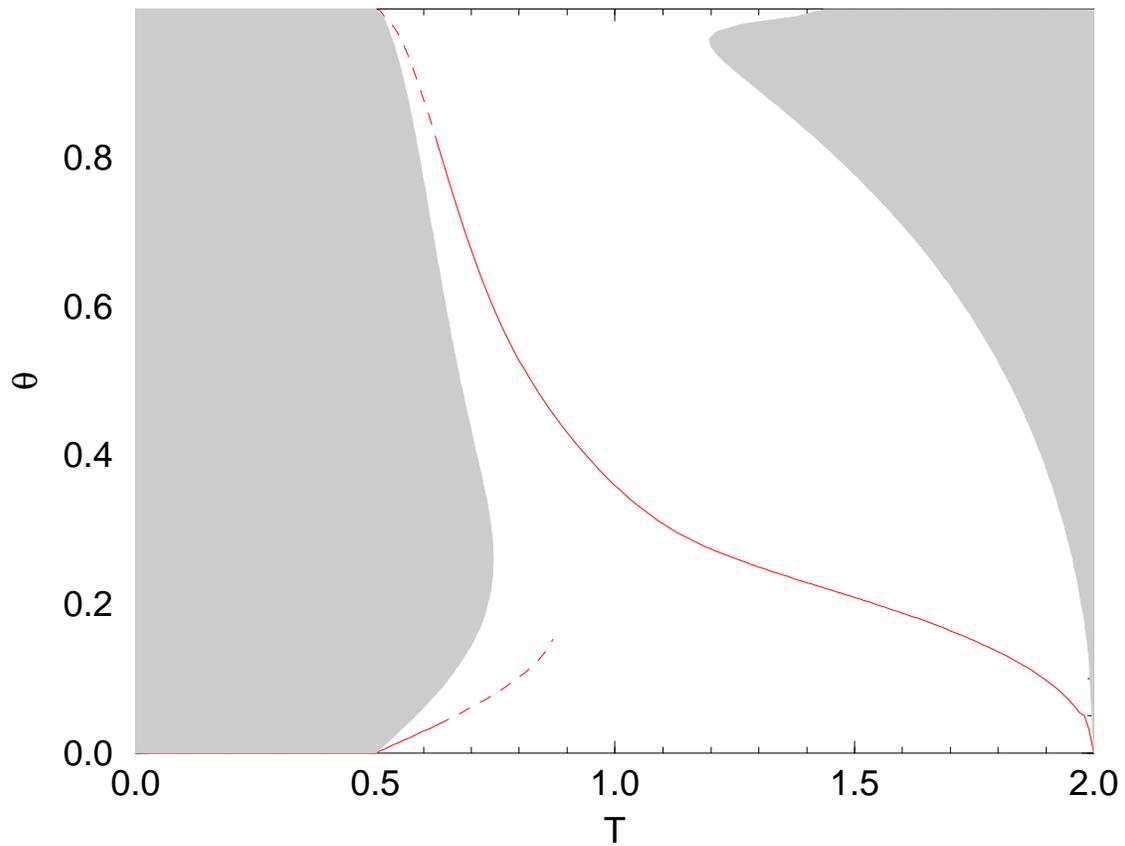,height=13cm}}
\caption{
The freezing threshold $\theta$ for the best overlap 
as a function of temperature $T$ 
for $p=2$ and $j_0=J=1$.
}
\label{bit4.vgr}
\end{figure}

We have also studied the dependence of the overlap
on varying the freezing threshold $\theta$
rather than the freezing fraction $f$.
However, Fig. \ref{bit4.vgr} shows that the optimal value of $\theta$
has a much larger dependence on the temperature.
This is due to the sensitive dependence
of the thermal averages of the spins on temperature.
At high temperatures, most spins are thermally agitated,
and the freezing threshold has to be set to a very low value
in order to freeze a given fraction of spins.
On the other hand, at low temperatures,
most spins are relatively stable, 
and the freezing threshold has to be set to a very high value
in order to keep a given fraction of spins dynamic in the second stage.
We conclude that the freezing fraction
is a better controlling parameter for the decoding performance.

\begin{figure}
\centering
\centerline{\psfig{figure=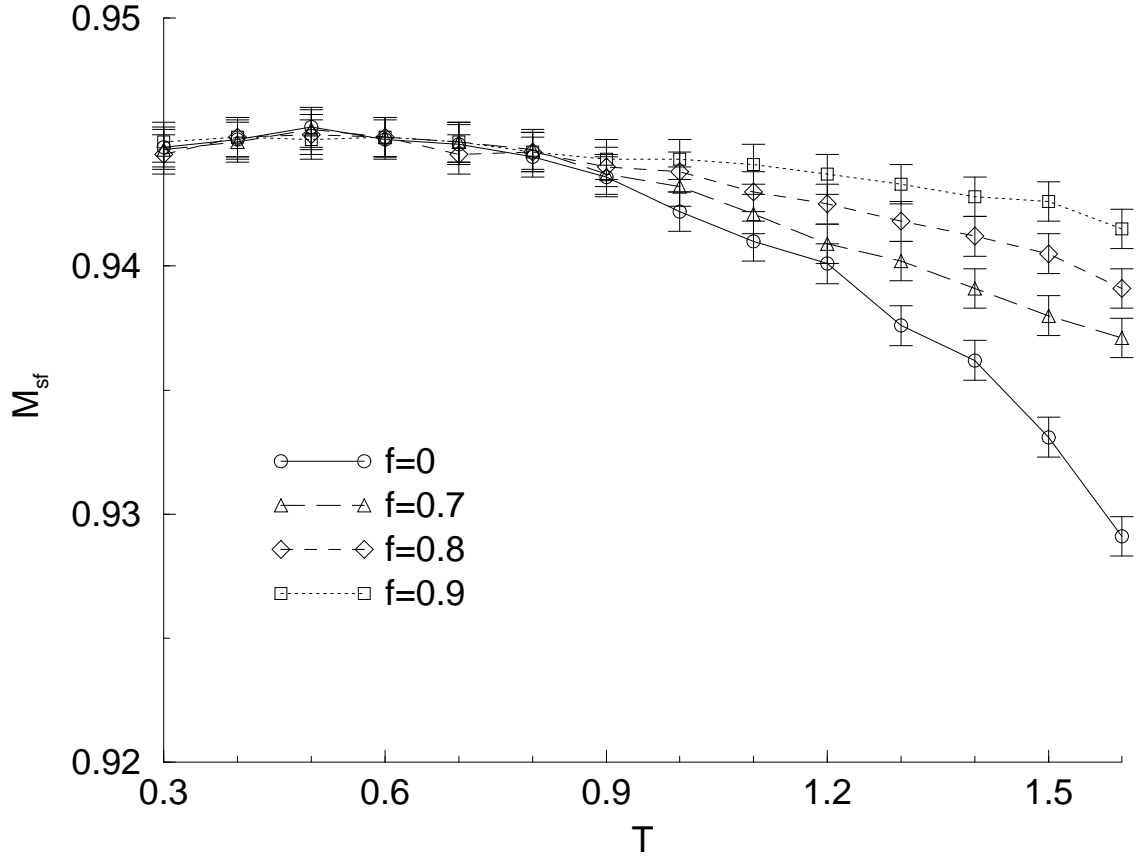,height=13cm}}
\caption{
Results of Monte Carlo simulations for the overlaps of selective freezing 
compared with that of the one-stage dynamics for $p=2$ and $j_0=J=1$,
corresponding to Fig. \ref{bif.vgr}.
The simulation parameters are: $N=1000$ with an initial overlap of 0.9 
and 200 samples. 
Each stage consists of 500 Monte Carlo steps per node for equilibration 
and 1000 Monte Carlo steps per node for averaging.
}
\label{sff2.vgr}
\end{figure}

The advantages of selective freezing are confirmed 
by Monte Carlo simulations shown in Fig. \ref{sff2.vgr}. 
For one-stage dynamics, the overlap is maximum at the Nishimori point 
($T_N=0.5$) as expected. 
However, it deterriorates rather rapidly 
when the decoding temperature increases. 
In contrast, selective freezing maintains a more steady performance, 
especially when $f=0.9$.

\section{THE MEAN-FIELD MODEL FOR IMAGE RESTORATION}

In conventional image restoration problems, 
a given degraded image consists of the set of pixels $\{\tau_i\}$, 
but not the set of exchange interactions $\{J_{i_1,\cdots,i_p}\}$. 
On the other hand, effective restoration requires 
the introduction of a model prior distribution of the pixels 
for smooth images. 
In this case the Hamiltonian corresponds 
to that of a random field Ising model,
\begin{equation}
	H\{\sigma\}=-h\sum_i\tau_i\sigma_i
	-{\beta_m\over z}\sum_{\langle ij\rangle}\sigma_i\sigma_j.
\end{equation}
In mean-field systems, each pixel $i$ has an extensive valency. 
The pixels $\tau_i$ are the degraded versions of the source pixels $\xi_i$, 
corrupted by noise which, for convenience, 
is assumed to be Gaussian with mean $a\xi_i$ and variance $\tau^2$, i.e.
\begin{equation}
        P(\tau_i|\xi_i)
        ={\exp\left[-{1\over 2\tau^2}(\tau_i-a\xi_i)^2\right]
        \over\sqrt{2\pi\tau^2}}.
\end{equation}
In turn, the source pixels satisfy
the prior distribution in Eq. (\ref{prior}).
Applying the cavity argument for mean-field systems,
the prior distribution becomes factorizable,
\begin{equation}
        P(\xi_i)={\exp(\beta_s m_0 \xi_i)
        \over 2\cosh\beta_sm_0},
\end{equation}
where $m_0=\tanh\beta_sm_0$.
The order parameter in the first stage is given by
\begin{equation}
        m\equiv{1\over N}\sum_i\langle\sigma_i\rangle
        ={1\over 2\cosh\beta_sm_0}\sum_{\xi=\pm 1}
        \exp(\beta_sm_0\xi)\int Dx\tanh U,
\end{equation}
where $U=\beta_mm+ha\xi+h\tau x$.
The overlap for the one-stage restoration process is given by
\begin{equation}
        M\equiv{1\over N}\sum_i\xi_i{\rm sgn}\langle\sigma_i\rangle
        ={1\over 2\cosh\beta_sm_0}\sum_{\xi=\pm 1}
        \exp(\beta_sm_0\xi)\xi
        {\rm erf}{\beta_mm+ha\xi\over\sqrt 2h\tau}.
\label{m_ca}
\end{equation}
Next we consider selective freezing in the second stage
with a freezing threshold $\theta$.
The freezing fraction is given by
\begin{equation}
        f\equiv{1\over N}\sum_i
        \Theta\left(\langle\sigma_i\rangle^2-\theta^2\right)
        ={1\over 2\cosh\beta_sm_0}\sum_{\xi=\pm 1}
        \exp(\beta_sm_0\xi)\left[1
        -{1\over 2}{\rm erf}{u_+(\xi)\over\sqrt 2}
        +{1\over 2}{\rm erf}{u_-(\xi)\over\sqrt 2}\right],
\end{equation}
where $u_\pm(\xi)=(\pm u_0-\beta_mm-ha\xi)/h\tau$
with $\tanh u_0=\theta$.
The order parameter of the second stage is given by
\begin{eqnarray}
        \tilde m
	&\equiv&{1\over N}\sum_i
        \left[\Theta(\theta^2-\langle\sigma_i\rangle^2)
        \langle\tilde\sigma_i\rangle
        +\Theta(\langle\sigma_i\rangle^2-\theta^2)
        {\rm sgn}\langle\sigma_i\rangle\right]\nonumber\\
        &=&{1\over 2\cosh\beta_sm_0}\sum_{\xi=\pm 1}
        \exp(\beta_sm_0\xi)\left[
        -{1\over 2}{\rm erf}{u_+(\xi)\over\sqrt 2}
        -{1\over 2}{\rm erf}{u_-(\xi)\over\sqrt 2}
        +\int_{u_-(\xi)}^{u_+(\xi)}Dx\tanh L\right],
\end{eqnarray}
where $L=\beta_m\tilde m+ha\xi+h\tau x$.
The overlap for selective freezing is given by
\begin{eqnarray}
        M_{\rm sf}
        &\equiv&{1\over N}\sum_i\xi_i
        \left[\Theta(\theta^2-\langle\sigma_i\rangle^2)
        {\rm sgn}\langle\tilde\sigma_i\rangle
        +\Theta(\langle\sigma_i\rangle^2-\theta^2)
        {\rm sgn}\langle\sigma_i\rangle\right]\nonumber\\
        &=&{1\over 2\cosh\beta_sm_0}\sum_{\xi=\pm 1}
        \exp(\beta_sm_0\xi)\xi{\rm erf}
        {g(\beta_m\tilde m)+ha\xi\over\sqrt 2h\tau},
\label{m_sf}
\end{eqnarray}
where
\begin{equation}
        g(\beta_m\tilde m)=\left\{\matrix{
        \beta_mm-u_0\hfill &&\hfill && \beta_m\tilde m<\beta_mm-u_0,\hfill\cr
        \beta_m\tilde m\hfill &&\hfill && 
	\beta_mm-u_0<\beta_m\tilde m<\beta_mm+u_0,\hfill\cr
        \beta_mm+u_0\hfill &&\hfill &&
	\beta_m\tilde m>\beta_mm+u_0.\hfill}\right.
\label{gfunc}
\end{equation}
We note that since the spin-glass interaction is absent in this case,
there are no trans-susceptibility effects.
This is unlike the case of error-correcting codes,
in which $\chi_{tr}$ is nonzero when $J$ is nonzero.

The three cases of the function $g(\beta_m\tilde m)$ in Eq. (\ref{gfunc})
correspond to three situations.
When $\beta_m\tilde m<\beta_mm-u_0$,
all the dynamic spins in the second stage
have negative thermodynamic averages
and therefore take the value $-1$ in the two-stage restoration process.
This is equivalent to a one-stage restoration process
in which all spins with thermodynamic averages
above the threshold $+\theta$ are frozen to $+1$,
and to $-1$ otherwise.
Similarly, when $\beta_m\tilde m>\beta_mm+u_0$,
all the dynamic spins in the second stage
have positive thermodynamic averages.
Only when $\beta_mm-u_0<\beta_m\tilde m<\beta_mm+u_0$,
do we have the dynamic spins frozen to partly $+1$ and partly $-1$.

We can consider the condition for the optimal performance
$M_{\rm sf}$ of selective freezing.
For a given distribution of data and noise,
$g(\beta_m\tilde m)$ is the only adjustable parameter in Eq. (\ref{m_sf}),
playing the same role as the adjustable parameter $\beta_mm$
for one-stage dynamics in Eq. (\ref{m_ca}).
In the space of $h$ and $\beta_m$,
the performance is optimal along the line
$h/\beta_\tau=\beta_mm/\beta_sm_0$ for one-stage dynamics \cite{nishiwong}
($\beta_\tau=a/\tau^2$ for Gaussian noise).
Analogously, there exists a line of optimal performance
defined by $h/\beta_\tau=g(\beta_m\tilde m)/\beta_sm_0$
for selective freezing.

\begin{figure}
\centering
\centerline{\psfig{figure=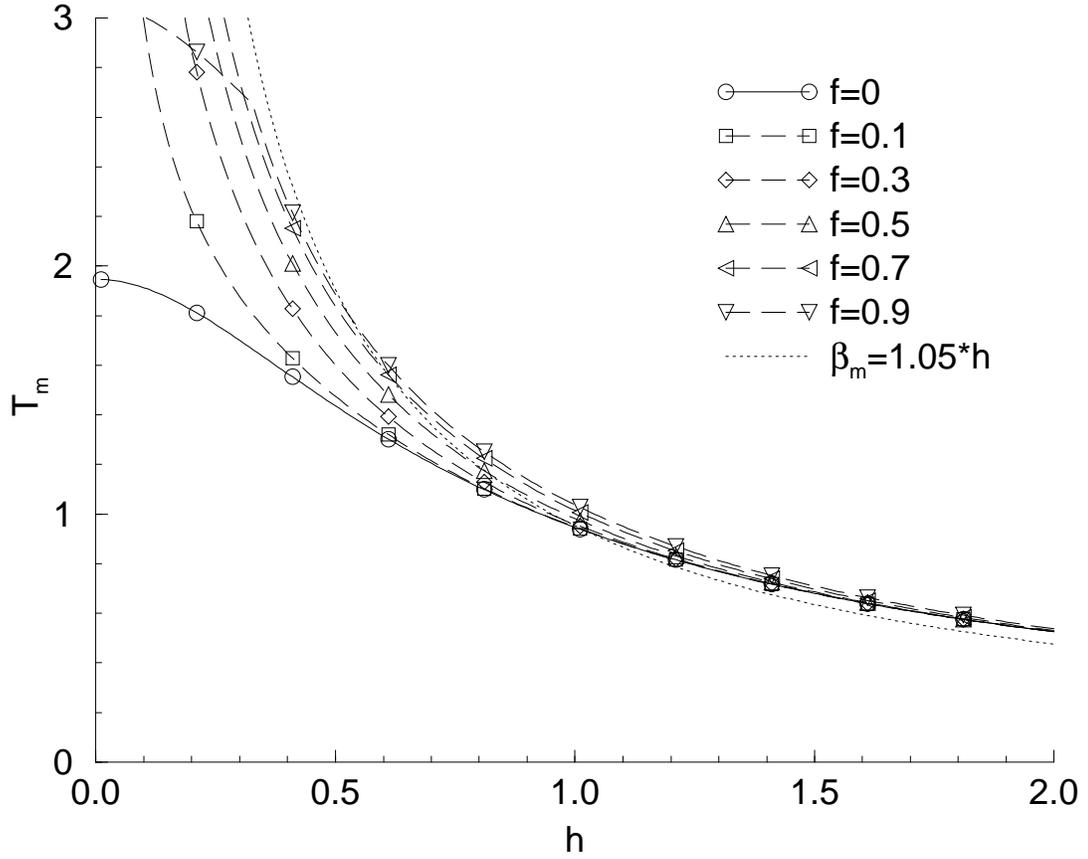,height=13cm}}
\caption{
The lines of optimal performance
in the space of the random-field strength $h$
and the restoration temperature $T_m\equiv\beta_m^{-1}$
in the mean-field model of image restoration
for $a=\tau=1$ and $\beta_s=1.05$.
The dotted line is the line of operation with $\beta_m/h$
set to the optimal ratio $\beta_s/\beta_\tau=1.05$.
At $f=0.9$ the dynamic spins are completely frozen to $+1$
to the left of the kink,
but only partially to the right.
}
\label{ilsf.vgr}
\end{figure}

An example of the lines of optimal performance is shown in Fig. \ref{ilsf.vgr}.
It is interesting to note the kinks for certain freezing fractions.
They correspond to transitions of cases in which
the dynamic spins are partially or completely frozen to $\pm 1$.

\begin{figure}
\centering
\centerline{\psfig{figure=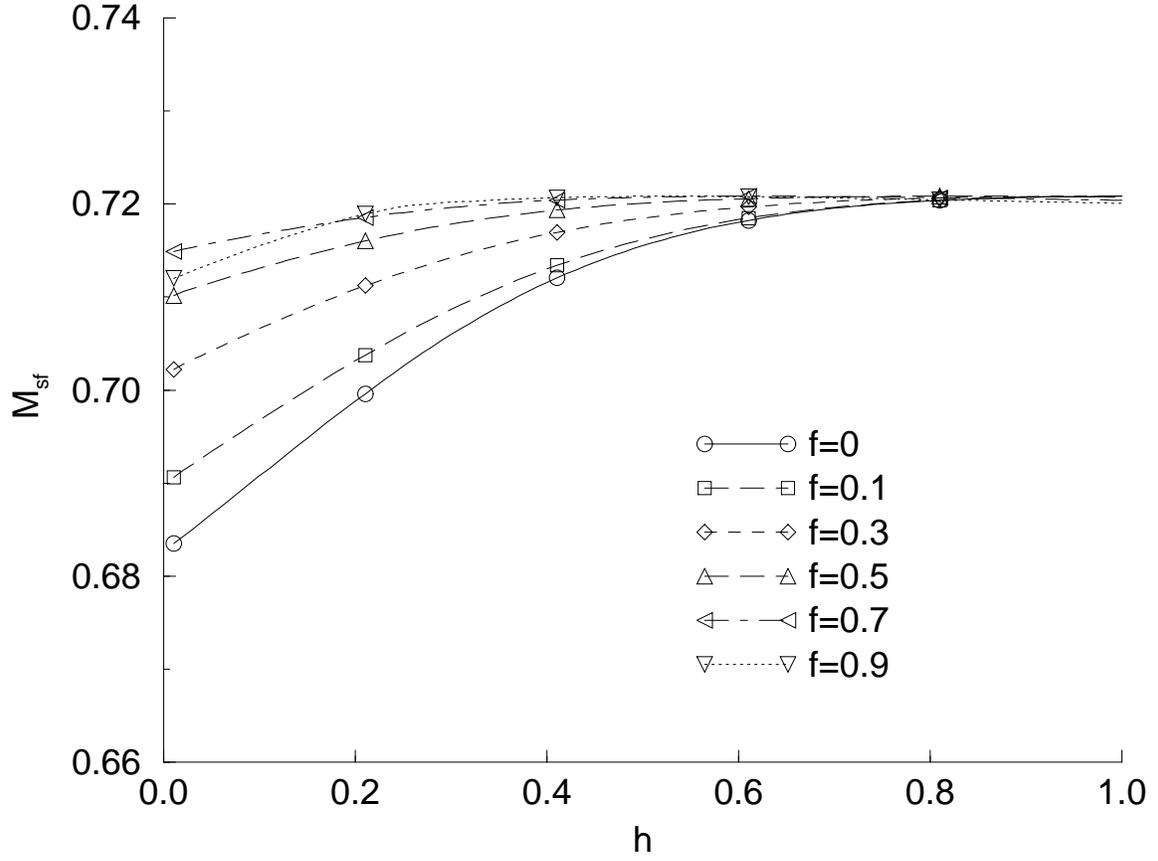,height=13cm}}
\caption{
The performance of selective freezing at $a=\tau=1$ and $\beta_s=1.05$,
with $\beta_m/h$ set to the optimal ratio $\beta_s/\beta_\tau=1.05$
for various freezing fraction $f$.
}
\label{irb10.vgr}
\end{figure}

A comparison of Eqs. (\ref{m_ca}) and (\ref{m_sf}) shows that
selective freezing performs as well as one-stage dynamics,
but cannot outperform it.
Nevertheless, selective freezing provides a rather stable performance
when the hyperparameters cannot be estimated precisely.
In image restoration, the usual practice
is to choose a fixed ratio of $\beta_m/h$.
Fig. \ref{irb10.vgr} confirms this stability along the line of operation
with $\beta_m/h$ set to the optimal ratio $\beta_s/\beta_\tau$.
Note especially that the lines with $f=0.7$ and $0.9$
attain a nearly optimal value of $M_{\rm sf}$
over a wide range of parameters.
The kink at $f=0.9$ is, again,
due to the appearance of the $-1$ frozen dynamic spins
(to the right of the kink).

The stable performance of selective freezing can be partly explained
by the proximity of the lines of optimal performance
with the line of operation which,
as discussed in \cite{nishiwong},
is an important factor in hyperparameter estimation.
This is illustrated by the optimal lines
for small values of $f$ near the Nishimori point
$(T_m,h)=(1.05^{-1},1)$ in Fig. \ref{ilsf.vgr}.

However, the advantage of selective freezing
does not only rely on the fortuitous combination of parameters.
Even when the parameters are not chosen optimally,
selective freezing still maintains a rather robust performance.
For example, along the line of optimal performance 
for $f=0.9$ in Fig. \ref{ilsf.vgr},
the bending at the kink only causes
a modest reduction in the overlap $M_{\rm sf}$ in Fig. \ref{irb10.vgr}.

To study the robustness of the performance of selective freezing,
we model a situation common in modern communication channels
carrying multimedia traffic,
which are often bursty in nature.
Since burstiness results in intermittent interferences,
we consider a noise with two Gaussian components,
each with its own characteristics.
A random fraction $f_1$ of the pixels
are influenced by Gaussian noise
with signal strength $a_1$ and noise variance $\tau_1^2$.
The rest of the pixels have strength $a_2$ and noise variance $\tau_2^2$.
Hence the distribution of the degraded pixels are
\begin{equation}
        P(\tau_i|\xi_i)
        =f_1{\exp\left[-{1\over 2\tau_1^2}(\tau_i-a_1\xi_i)^2\right]
        \over\sqrt{2\pi\tau_1^2}}
        +f_2{\exp\left[-{1\over 2\tau_2^2}(\tau_i-a_2\xi_i)^2\right]
        \over\sqrt{2\pi\tau_2^2}},
\end{equation}
where $f_2=1-f_1$.
The equations for the order parameters can be generalized
from the single component case in a straightforward manner.

\begin{figure}
\centering
\centerline{\psfig{figure=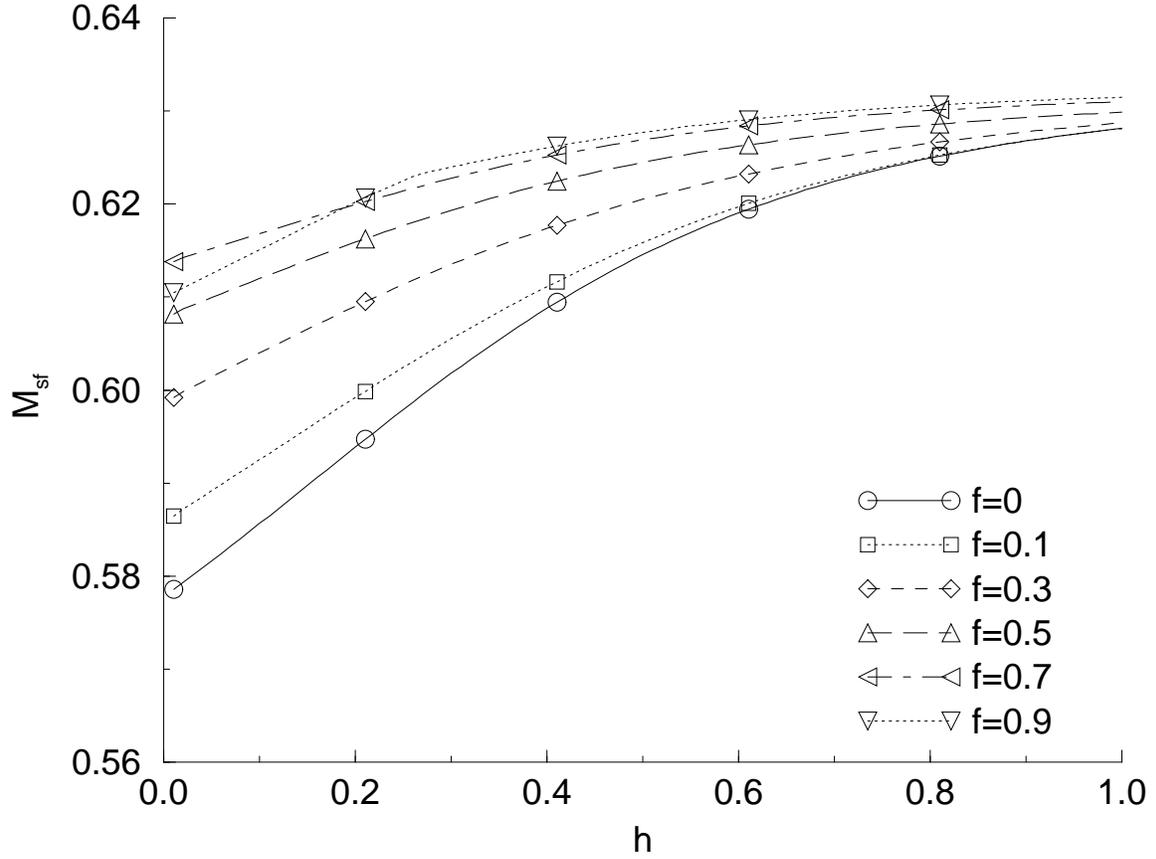,height=13cm}}
\caption{
The performance of selective freezing
with 2 components of Gaussian noise at $\beta_s=1.05$,
$f_1=4f_2=0.8$, $a_1=5a_2=1$ and $\tau_1=\tau_2=1$.
The restoration agent operates by assuming the majority component,
i.e. $\beta_m/h=\beta_s\tau_1^2/a_1$.
}
\label{ird8.vgr}
\end{figure}

A case of interest is that the restoration agent operates
on the assumption of the characteristics
of the majority component of the channel, say the first component.
Hence it operates at the ratio $\beta_m/h=\beta_s\tau_1^2/a_1$.
Suppose the Gaussian noise is partly interrupted
to take the characteristics of the second component,
but the operation parameters cannot be adjusted soon enough,
then there will be a degradation of the quality of the restored images.
In the example in Fig. \ref{ird8.vgr},
the reduction of the overlap $M_{\rm sf}$ for selective freezing
is much more modest than the one-stage process ($f=0$).

\begin{figure}
\centering
\centerline{\psfig{figure=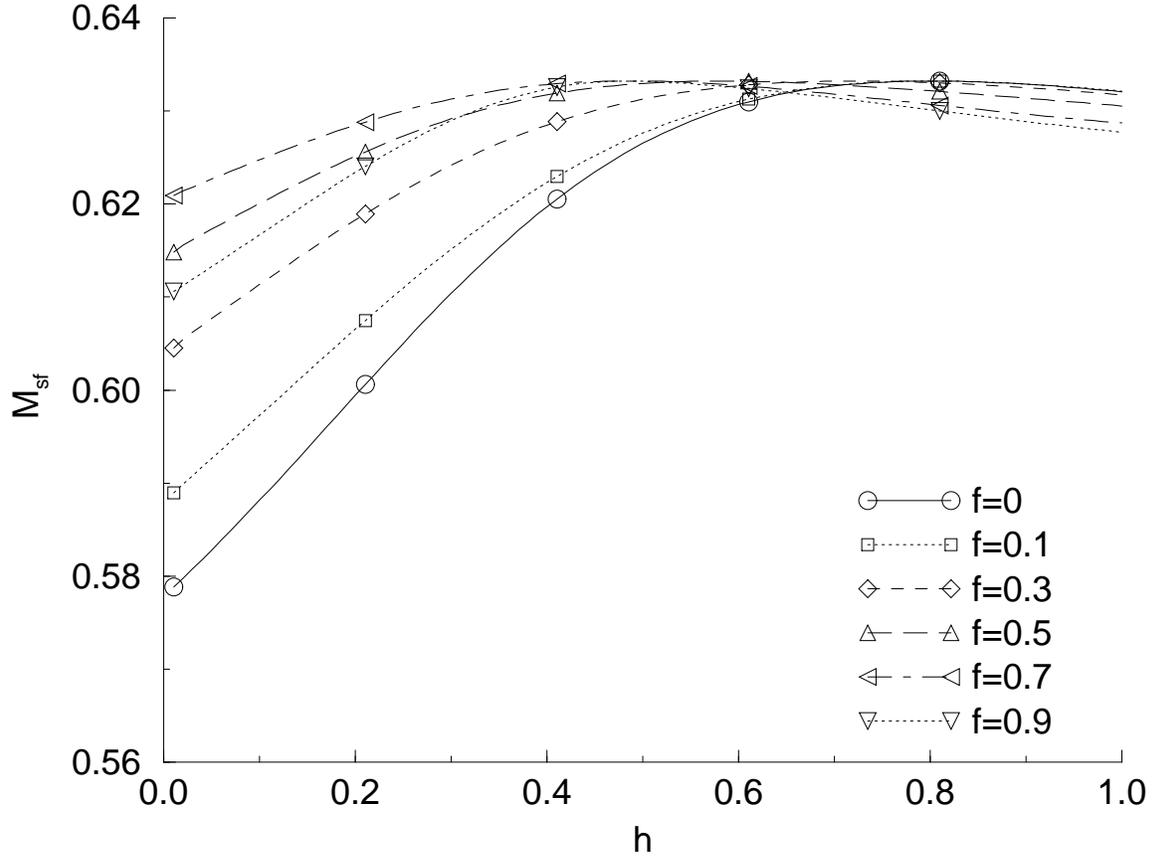,height=13cm}}
\caption{
Same as Fig. \ref{ird8.vgr}, except that the restoration agent
operates with the ratio $\beta_m/h=\beta_s^*\tau^{*2}/a^*$,
where $\beta_s^*$, $\tau^*$ and $a^*$
are estimated from Eq. (\ref{estimate1})-(\ref{estimate3}).
}
\label{irc8.vgr}
\end{figure}

An alternative situation is that the restoration agent is able
to detect the changes in the average signal strengths and noise variance,
but still operates on the assumption
of a single-component Gaussian channel.
Suppose that such simple statistics as
$\langle{\rm sgn}\tau_i\rangle$, $\langle\tau_i\rangle$
and $\langle\tau_i^2\rangle$ are accessible.
Then the parameters $m_0^*$, $a^*$ and $\tau^*$
estimated by the restoration agent are obtained,
for $\tau_1=\tau_2=\tau$, from the solutions of
\begin{eqnarray}
        m_0^*{\rm erf}{a^*\over\sqrt 2\tau^*}
        &=&\langle{\rm sgn}\tau_i\rangle=m_0\left[
        f_1{\rm erf}{a_1\over\sqrt 2\tau_1}+
        f_2{\rm erf}{a_2\over\sqrt 2\tau_2}\right],
\label{estimate1}\\
        m_0^*a^*
        &=&\langle\tau_i\rangle=m_0\left[f_1a_1+f_2a_2\right],
\label{estimate2}\\
        a^{*2}+\tau^{*2}
        &=&\langle\tau_i^2\rangle
        =f_1(a_1^2+\tau_1^2)+f_2(a_2^2+\tau_2^2),
\label{estimate3}
\end{eqnarray}
and $\beta_s^*=\tanh^{-1}m_0^*/m_0^*$.
Using these estimated parameters,
the performances in Fig. \ref{irc8.vgr} improve over their counterparts
based on only the majority component in Fig. \ref{ird8.vgr}.
Still, one-stage restoration cannot avoid the performance drop
when $h$ vanishes, whereas correspondingly,
selective freezing has a much more gentle drop in performance.

\begin{figure}
\centering
\centerline{\psfig{figure=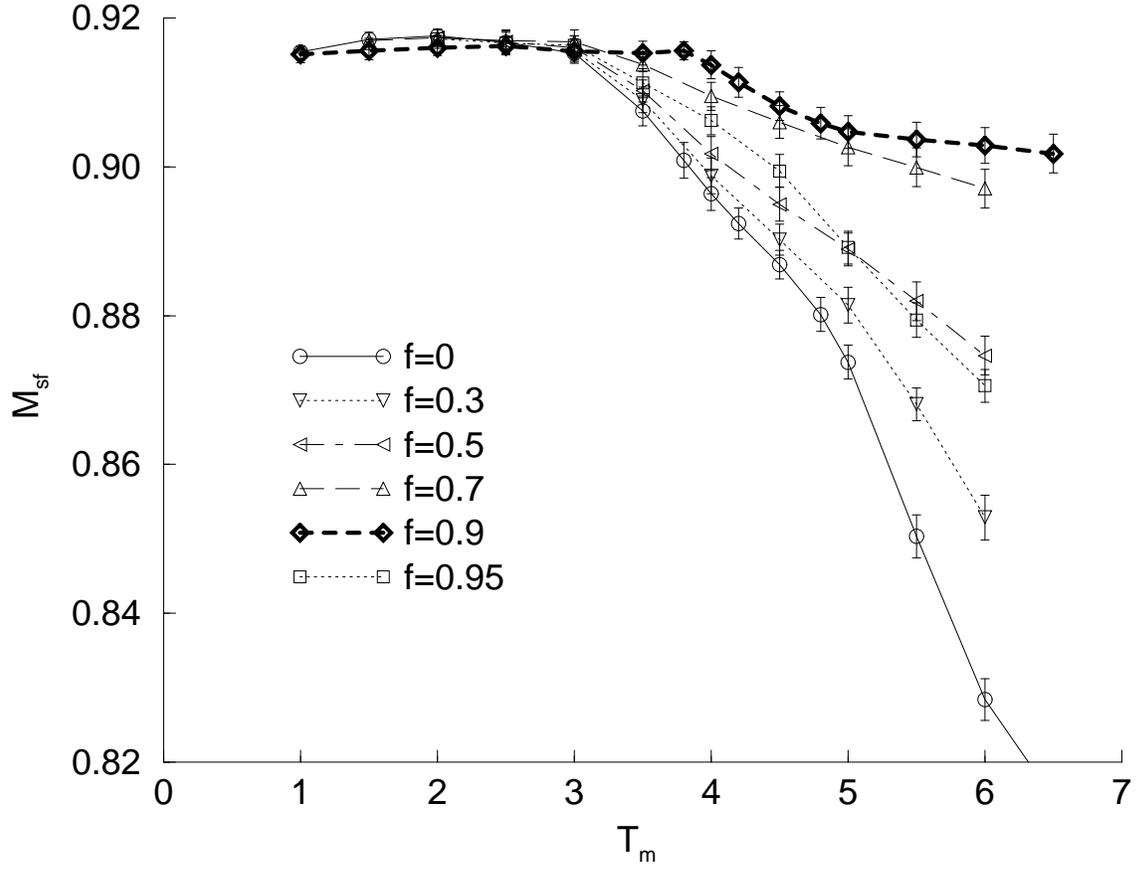,height=13cm}}
\caption{
Results of Monte Carlo simulations for the overlaps of selective freezing 
compared with that of the one-stage dynamics for two-dimensional images 
generated at the source prior temperature $T_s=2.15$. 
The simulation parameters are: $N=50\times 50$, 
with an initial overlap of 0.8 and 1000 samples. 
Each stage consist of 
1000 Monte Carlo steps per node for averaging.
}
\label{imsf.eps}
\end{figure}

It is interesting to study 
the more realistic case of two-dimensional images, 
since we have so far presented analytical results 
for the mean field model only. 
As confirmed by the results for Monte carlo simulations 
in Fig. \ref{imsf.eps}, 
the overlaps of selective freezing are much more steadier 
than that of the one-stage dynamics 
when the decoding temperature changes. 
This steadiness is most remarkable for a freezing fraction of $f=0.9$.

\section{DISCUSSIONS}

We have introduced a multistage technique
for error-correcting codes and image restoration,
in which the information extracted from the former stage
can be used selectively to improve the performance of the latter one.
While the overlap $M_{\rm sf}$ of the selective freezing
is bounded by the optimal performance of the one-stage dynamics
derived in \cite{nishiwong},
it has the advantage of being tolerant
to uncertainties in hyperparameter estimation.
The performance is especially steady
when the fraction of frozen spins,
rather than the threshold of their thermodynamic averages,
is fixed in the process.
This is confirmed by both analytical and simulational results 
for mean-field and finite-dimensional models.
As an example, we have illustrated its advantage of robustness
when the noise distribution is composed of more than one Gaussian components,
such as in the case of modern communication channels
supporting multimedia applications.

We found that selective freezing is most useful
when more than one hyperparameters have to be estimated,
as illustrated by the example of image restoration,
where both $\beta_m$ and $h$ have to be estimated.
In the example of error-correcting codes discussed in Section III,
there is only one hyperparameter $T_m$,
and it is found that selective freezing has performance advantages
only when $T_m$ is chosen above the Nishimori point.
However, more than one hyperparameters
are often present in practical applications.

Selective freezing can be generalized to more than two stages,
in which spins that remain relatively stable in one stage
are progressively frozen in the following one.
It is expected that the performance can be even more robust.

While the multistage process described here has a robust performance,
it does not raise the critical temperature
or the critical noise level
for the existence of the ordered phase.
Nor can it widen the basin of attraction for the ordered phase.
Other multistage processes, proposed in \cite{wong} for neural networks,
may be able to achieve this.
This remains an area for further research.

We have made progress in the theoretical treatment
of multistage processes using the cavity method.
It allows the thermal averages of spins
to be expressed in terms of the cavity fields.
Since a cavity field is uncorrelated with the spin in consideration,
it can in turn be expressed in terms of
the means and covariances of the spin averages,
thereby arriving at a set of self-consistent equations
for the order parameters.
In particular, there appears a trans-susceptibility term
since variations of the cavity field in the first stage
are correlated with the spin average in the second stage
due to the selective nature of the freezing process in the second stage.
However, for the ordered phase considered in this paper,
the effects of the trans-susceptibility term is not too large
except near the phase boundary.

On the other hand, 
we have a remark about the basic assumption of the cavity method, 
namely that the addition or removal of a spin 
causes a small change in the system 
describable by a perturbative approach. 
In fact, adding or removing a spin may cause 
the thermal averages of other spins 
to change from below to above the thresholds $\pm\theta$ (or vice versa). 
This change, though often small, 
induces a non-negligible change of the thermal averages 
from fractional values to the frozen values of $\pm 1$ (or vice versa) 
in the second stage.
The perturbative analysis of these changes is only approximate. 
The situation is reminiscent of similar instabilities 
in other disordered systems such as the perceptron, 
and are equivalent to Almeida-Thouless instabilities 
in the replica method \cite{wong3}. 
A full treatment of the problem would require 
the introduction of a rough energy landscape \cite{wong3}, 
or the replica symmetry breaking ansatz in the replica method \cite{mezard}. 
Nevertheless, previous experiences on disordered systems 
showed that the corrections made by a more complete treatment 
may not be too large in the ordered phase.
For example, corresponding analytical and simulational results 
in Figs. \ref{bif.vgr} and \ref{sff2.vgr} respectively 
are close to each other.

In practical implementations of error-correcting codes, 
algorithms based on belief-propagation methods, 
rather than Monte Carlo methods, are often employed \cite{frey}. 
It has recently been shown that such decoded messages 
converge to the solutions of the TAP equations 
in the corresponding thermodynamic system \cite{kabasaad2}. 
Again, the performance of these algorithms are sensitive 
to the estimation of hyperparameters. 
We propose that the selective freezing procedure 
has the potential to make these algorithms more robust.

Incidentally, multistage dynamics has also been applied
in the recently popular turbo codes \cite{turbo}.
Messages are coded in sequences with two possible permutations
and at each iterative stage,
the information derived from decoding one sequence
is fed to the other in the form of external fields for each bit.
The techniques developed in the present context
can be used to study this iterative process.

\acknowledgments
KYMW wishes to thank Tokyo Institute of Technology for hospitality.
HN is grateful to Hong Kong University of Science and Technology
for hospitality.
This work was partially supported by research grant HKUST6157/99P
of the Research Grant Council of Hong Kong.

\appendix
\section{Thermal averages of spins}

In this appendix we derive Eq. (\ref{cavity})
starting from the clustering property Eq. (\ref{cluster}). 
For convenience we illustrate the derivation for $p=2$.
We separate the Hamiltonian into two parts,
one does not contain $\sigma_1$ and the other does.
Hence
\begin{equation}
        H=H^{\backslash 1}-\beta\sum_{j>1}J_{1j}\sigma_1\sigma_j.
\end{equation}
Thus the thermal average can be written as
\begin{equation}
        \langle\sigma_1\rangle={
        {\rm Tr}^{\backslash 1}e^{-H^{\backslash 1}}
        {\rm Tr}_1\sigma_1
        \exp\left(\beta\sigma_1\sum_jJ_{1j}\sigma_j\right)
        /{\rm Tr}^{\backslash 1}e^{-H^{\backslash 1}}\over
        {\rm Tr}^{\backslash 1}e^{-H^{\backslash 1}}
        {\rm Tr}_1
        \exp\left(\beta\sigma_1\sum_jJ_{1j}\sigma_j\right)
        /{\rm Tr}^{\backslash 1}e^{-H^{\backslash 1}}}.
\label{thermal}
\end{equation}
Expanding the exponential function in the denominator
and tracing over $\sigma$, we get
\begin{equation}
        {\rm Den.}=
        2\sum_{n\ \rm even}{\beta^n\over n!}
        \sum_{j_1\cdots j_n}J_{1j_1}\cdots J_{1j_n}
        \langle\sigma_{j_1}\cdots\sigma_{j_n}\rangle^{\backslash 1}.
\end{equation}
Next, we use the clustering property to factorize the thermal average
$\langle\sigma_{j_1}\cdots\sigma_{j_n}\rangle^{\backslash 1}$.
For the coupling distribution specified by Eq. (\ref{jdis}),
only two kinds of contributions are significant
in the summation over the indices $j_1\cdots j_n$.
In the first kind, an index $j$ remains distinct from the rest,
contributing a factor of $J_{1j}\langle\sigma_j\rangle^{\backslash 1}$.
In the second kind, two indices become paired up.
However, when $j$ and $k$ pair up,
the thermal average $\langle\sigma_j\sigma_k\rangle^{\backslash 1}$
becomes 1 instead of $(\langle\sigma_j\rangle^{\backslash 1})^2$.
Hence the additional contribution due to the pairing is
$J_{1j}^2[1-(\langle\sigma_j\rangle^{\backslash 1})^2]$.
Other than these, the contributions
due to the pairing of three or more indices
are smaller by factors of $N$.

The denominator can now be considered a summation over $n$ and $m$,
which are respectively the total number of indices
and the number of pairs of paired indices appearing in a term.
The number of such terms is $n!/m!2^m(n-2m)!$. Hence
\begin{equation}
        {\rm Den.}=
        2\sum_{n\ \rm even}\sum_{m=0}^{n/2}
        {\beta^n\over n!}{n!\over m!2^m(n-2m)!}
        \left[\sum_jJ_{1j}\langle\sigma_j\rangle^{\backslash 1}\right]^{n-2m}
        \left\{\sum_jJ_{1j}^2\left[1-
        (\langle\sigma_j\rangle^{\backslash 1})^2\right]\right\}^m,
\end{equation}
which can be simplified to
\begin{equation}
        {\rm Den.}=
        2\exp\left\{{1\over 2}\beta^2\sum_j
        J_{1j}^2\left[1-(\langle\sigma_j\rangle^{\backslash 1})^2
        \right]\right\}\cosh\left\{\beta\sum_j
        J_{1j}\langle\sigma_j\rangle^{\backslash 1}\right\}.
\label{den}
\end{equation}
Similarly, the numerator can be written as
\begin{equation}
        {\rm Num.}=
        2\exp\left\{{1\over 2}\beta^2\sum_j
        J_{1j}^2\left[1-(\langle\sigma_j\rangle^{\backslash 1})^2
        \right]\right\}\sinh\left\{\beta\sum_j
        J_{1j}\langle\sigma_j\rangle^{\backslash 1}\right\}.
\label{num}
\end{equation}
Substituting Eq. (\ref{den}) and (\ref{num}) into Eq. (\ref{thermal}),
we arrive at Eq. (\ref{cavity}).

\section{Change in thermal averages on removal of a spin}

In this appendix we derive Eq. (\ref{2-change}).
For convenience we illustrate the derivation for $p=2$.
We separate the Hamiltonian into four parts:
(a) does not contain spins 1 and $j$,
(b) contains only spins 1 and $j$,
(c) contains spin 1 but not $j$,
(d) contains spin $j$ but not 1.
This yields
\begin{equation}
        H=H^{\backslash 1j}
        -\beta J_{1j}\sigma_1\sigma_j
        -\beta\sum_{k\ne 1j}J_{k1}\sigma_k\sigma_1
        -\beta\sum_{k\ne 1j}J_{kj}\sigma_k\sigma_j.
\end{equation}
The thermal average of $\sigma_j$ can then be written as
\begin{equation}
        \langle\sigma_j\rangle={
        {\rm Tr}_{1j}
        {\rm Tr}^{\backslash 1j}e^{-H}\sigma_j/
        {\rm Tr}^{\backslash 1j}e^{-H^{\backslash 1j}}\over
        {\rm Tr}_{1j}
        {\rm Tr}^{\backslash 1j}e^{-H}/
        {\rm Tr}^{\backslash 1j}e^{-H^{\backslash 1j}}}.
\end{equation}
Using the mean-field technique developed in Appendix A,
the denominator can be written as
\begin{eqnarray}
	{\rm Den.}=
	{\rm Tr}_{1j}\exp\biggl\{
	\beta J_{1j}\sigma_1\sigma_j
	&+&\beta\sum_{k\ne 1j}\langle\sigma_k\rangle^{\backslash 1j}
	(J_{k1}\sigma_1+J_{kj}\sigma_j)\nonumber\\
	&+&{1\over 2}\beta^2\sum_{k\ne 1j}
	\left[1-(\langle\sigma_k\rangle^{\backslash 1j})^2\right]
	(J_{k1}\sigma_1+J_{kj}\sigma_j)^2\biggr\}.
\end{eqnarray}
After collecting terms and discarding negligible ones,
\begin{equation}
	{\rm Den.}=
	{\rm Tr}_{1j}\exp\left\{
	\beta\sigma_1\sum_{k\ne 1j}J_{1k}
	\langle\sigma_k\rangle^{\backslash 1j}+
	\beta\sigma_j\sum_{k\ne 1j}J_{jk}
	\langle\sigma_k\rangle^{\backslash 1j}
	+\beta J_{1j}\sigma_1\sigma_j
	+\beta^2(1-q)J^2\right\}.
\end{equation}
Together with a similar manipulation of the numerator, we obtain
\begin{equation}
        \langle\sigma_j\rangle=
        \tanh\beta\left(h_j^{\backslash 1}
        +J_{j1}\tanh\beta h_1^{\backslash j}\right),
\end{equation}
whose Taylor expansion yields
\begin{equation}
        \langle\sigma_j\rangle=
        \langle\sigma_j\rangle^{\backslash 1}
        +\left(\beta{\rm sech}^2\beta h_j^{\backslash 1}\right)
	\left(J_{j1}\tanh\beta h_1^{\backslash j}\right),
\end{equation}
which becomes Eq. (\ref{2-change}) for the case $p=2$.



\end{document}